# Contrastive Learning of Shared Spatiotemporal EEG Representations Across Individuals for Naturalistic Neuroscience

Xinke Shen[a,b,#], Lingyi Tao[c,#], Xuyang Chen[d], Sen Song[a,e], Quanying Liu[b,*], Dan Zhang[c,e,*]

[a]Department of Biomedical Engineering, Tsinghua University, Beijing, China

[b]Department of Biomedical Engineering, Southern University of Science and Technology, Shenzhen, China

[c]Department of Psychological and Cognitive Sciences, Tsinghua University, Beijing, China

[d]Department of Biology, Southern University of Science and Technology, Shenzhen, China

[e]Tsinghua Laboratory of Brain and Intelligence, Tsinghua University, Beijing, China

[#]These authors contributed equally to this work.

*Correspondence: liuqy@sustech.edu.cn; dzhang@tsinghua.edu.cn

**Abstract**: Neural representations induced by naturalistic stimuli offer insights into how humans respond to stimuli in daily life. Understanding neural mechanisms underlying naturalistic stimuli processing hinges on the precise identification and extraction of the shared neural patterns that are consistently present across individuals. Targeting the Electroencephalogram (EEG) technique, known for its rich spatial and temporal information, this study presents a framework for Contrastive Learning of Shared SpatioTemporal EEG Representations across individuals (CL-SSTER). CL-SSTER utilizes contrastive learning to maximize the similarity of EEG representations across individuals for identical stimuli, contrasting with those for varied stimuli. The network employed spatial and temporal convolutions to simultaneously learn the spatial and temporal patterns inherent in EEG. The versatility of CL-SSTER was demonstrated on three EEG datasets, including a synthetic dataset, a natural speech comprehension EEG dataset, and an emotional video watching EEG dataset. CL-SSTER attained the highest inter-subject correlation (ISC) values compared to the state-of-the-art ISC methods. The latent representations generated by CL-SSTER exhibited reliable spatiotemporal EEG patterns, which can be explained by properties of the naturalistic stimuli. CL-SSTER serves as an interpretable and scalable framework for the identification of inter-subject shared neural representations in naturalistic neuroscience.

# 1. Introduction

Neuroscience research is experiencing a paradigm shift towards naturalistic environments, emphasizing the use of stimuli (such as narratives and movies) that are high-dimensional, rich in information, and closely aligned with what we encounter in daily life (Jääskeläinen et al., 2021; Nastase et al., 2019; Sonkusare et al., 2019). The continuous and dynamic nature of naturalistic stimuli presents significant challenges to traditional event-related analysis methods in neuroscience, which are primarily designed for parametric experiments using isolated, controlled stimuli (Ben-Yakov et al., 2012; Sonkusare et al., 2019). This has spurred the development of data-driven methods like inter-subject correlation (ISC), which revealed the shared response to naturalistic stimuli across subjects (Hasson et al., 2004; Nastase et al., 2019; Zhang, 2018). ISC analyses have provided novel insights into neural substrates of speech comprehension (Li et al., 2021; Li & Zhang, 2024; Silbert et al., 2014; Stephens et al., 2010), attention (Chen et al., 2023c; Ki et al., 2016), emotion (Ding et al., 2021; Dmochowski et al., 2014; Nummenmaa et al., 2012), and social cognition (Chen et al., 2023a; Hakim et al., 2023; Liu et al., 2021; Wallace, 2023). For instance, ISC uncovered the cortical temporal hierarchy extending from primary sensory to heteromodal areas (Hasson et al., 2008).

Additionally, ISC from a small cohort of participants can predict wider population preferences for the stimuli (Dmochowski et al., 2014). However, typical ISC methods hinge on linear correlations or coherence between vertices or channels (Balconi & Angioletti, 2023; Balconi & Vanutelli, 2018; Coomans et al., 2021; Cui et al., 2012; Filho et al., 2016; Hakim et al., 2023; Hasson et al., 2004; Holper et al., 2012; Jiang et al., 2015; Nastase et al., 2019; Richard et al., 2021; Stevens & Galloway, 2022; Xu et al., 2024), overlooking potential latent generators of brain activities that are not readily apparent in the observational data. This has prompted the development of methods that characterize ISC in latent space.

Identifying reliable latent factors of neural activities has been a longstanding aim of neuroscience studies, as high-dimensional neural signals can often be explained by latent biophysically meaningful factors (Churchland et al., 2012; Pandarinath et al., 2018; Shine et al., 2019; Yu et al., 2009). In ISC analyses, "hyperalignment" methods transform the functional Magnetic Resonance Imaging (fMRI) data from a high-dimensional voxel space to a common voxel space, aligning activation patterns across individuals for identical states (Haxby et al., 2020). Similarly, correlated component analysis (CorrCA) applies a linear spatial projection to EEG data, enhancing ISC in latent space (Dmochowski et al., 2012; Parra et al., 2019). These techniques uncover shared neural representations more effectively than traditional ISC measures. However, their reliance on linear transformations overlooks the complex interactions in high-dimensional neural signals. For instance, CorrCA does not consider temporal or frequency aspects of EEG data, which is crucial for interpreting responses to naturalistic stimuli (Cohen, 2017; Kaneshiro et al., 2020). Consequently, these methods often yield modest ISC values in EEG data, even in latent space (e.g., ISC < 0.05, Parra et al., 2019). The joint temporal and spatial EEG representations underlying naturalistic stimuli still lacks full investigation.

Deep neural networks, empowered by the emerging contrastive learning technique, enable flexible and effective representation learning from neural signals (Katthi & Ganapathy, 2021a, 2021b; Pandarinath et al., 2018; Schneider et al., 2023). Contrastive learning identifies latent representations through the contrast of positive and negative sample pairs (Jaiswal et al., 2020; Liu et al., 2021). It has attained meaningful and consistent neural representations in animal studies (Schneider et al., 2023) and demonstrated promising performance in speech decoding (Schneider et al., 2023), visual decoding (Chen et al., 2023a; Chen et al., 2023b), sleep staging (Mohsenvand et al., 2020), emotion recognition (Shen et al., 2023) and mental disorder identification (Aglinskas et al., 2022; Tong et al., 2024) from neural signals. Moreover, it has been further proposed as a general associative learning approach for both biological and artificial intelligence (Vong et al., 2024). Leveraging the capabilities of contrastive learning to extract inter-subject shared neural representations could offer new breakthroughs in naturalistic neuroscience research.

In this study, we introduce a Contrastive Learning approach to identify Shared SpatioTemporal EEG Representations (CL-SSTER) across individuals. Our primary goal is to extract shared neural representations among subjects more effectively, which can be indicated by increased ISC values. Additionally, we aim to ensure the interpretability of our model to explore the biophysiological significance of the shared representations. In contrastive learning, we maximize the model's discriminative power of which EEG sample pair is in response to the same stimuli and which pair is not (Fig. 1a, b). A two-layered convolutional neural network is leveraged to learn temporal and spatial filters of EEG, respectively (Fig. 1c, Fig. S1). Building upon our prior research on contrastive learning methods for cross-subject emotion recognition (Shen et al., 2023), this study systematically investigates the model's design and demonstrates it as a robust tool in uncovering shared patterns across subjects for naturalistic neuroscience. We validate the method on three datasets: a synthetic EEG dataset, a natural speech comprehension EEG dataset published by Broderick et al. (2019) and an emotional video watching EEG dataset (FACED, Chen et al., 2023b). The method is compared with the state-of-the-art CorrCA method and correlations in sensor space. ISC values increase from 0.010 in sensor space to 0.424 on average for the synthetic dataset and from 0.034 to 0.062 on average across the two real-world datasets. The ISC values identified by the CL-SSTER method also exceed those identified by CorrCA method by 37.4% on average across the real-world datasets. Furthermore, we visualize the inter-subject shared EEG spatiotemporal representations obtained by CL-SSTER and explore their relationships with stimuli information (i.e., speech amplitude envelope, semantic dissimilarity, video brightness and emotional ratings) to elucidate their biophysiological meanings. The results not only align with but also extend the classical findings about naturalistic speech and visual processing. The proposed CL-SSTER method is made publicly available as a Python toolbox, encouraging further research and application in the field.

## 2. Materials and methods

*2.1 The datasets*

*2.1.1   The synthetic dataset*

To evaluate whether the method can identify ground-truth inter-subject shared representations correctly, we constructed a synthetic EEG dataset with inter-subject shared response and subject-specific response/noise. The synthetic EEG comprised three components: the target inter-subject shared responses (P1), subject-specific responses (P2) originating from the same brain region as the shared responses, and subject-specific responses (P3) from different brain regions (Fig. 2). The P1 component was an even combination of randomly generated broadband pink noise and pink noise filtered within a chosen frequency range (see Fig. 3a for the frequency responses of the P1 component), and was replicated across subjects. P2 and P3 were randomly generated broadband pink noise specific to each subject. We assume that P1 and P2 were from a common single source, so they are both projected to the sensor space by a random matrix $A_1 \in \mathbb{R}^{1 \times M}$, where $M$ is the number of sensors. P3 was constructed as random activities from various sources heterogeneous from P1 and P2, and was projected to the sensor space by a random matrix $A_2 \in \mathbb{R}^{M \times M}$. P1 and P2 were combined at a ratio of 0.24:0.76 (SNR = -10 dB) before projection, and then combined in the sensor space with P3 at a ratio of 0.15:0.85 (SNR = -15 dB). The combination ratio was so set as to align the maximal ISC level between the synthetic data (ISC = 0.010±0.000) and the Broderick dataset (ISC = 0.012±0.000). Other parameters also approximate the setting of the Broderick dataset (number of trials = 20, length of each trial = 150 seconds, number of subjects = 20, number of EEG sensors = 128, EEG sampling rate =128 Hz).

We varied the frequency response of the target signal (P1) in order to evaluate model performance more comprehensively. We selected four physiologically meaningful frequency ranges deviant from conventional EEG bands: 2-8 Hz (speech-related, e.g. Bröhl & Kayser, 2021), 8-10 Hz (low alpha, e.g. Bazanova & Vernon, 2014), 14-20 Hz (low beta, e.g. Abhang et al., 2016), and 25-35 Hz (emotion -related, e.g., Zheng et al., 2015).

*2.1.2    The Broderick dataset of natural speech comprehension*

In the Broderick dataset (Broderick et al., 2019), 128-channel EEG recordings were collected from 19 subjects as they listened to an audiobook. The audiobook was a popular American fiction written in mid-20th century. The experimental protocol consisted of 20 trials per subject, each trial lasting slightly under 180 seconds. The trials adhered to the original timeline of the audiobook, without any repetitions or discontinuities. The publicly available data have been downsampled from 512 Hz to 128 Hz and segmented into trials. For additional details on the dataset, please refer to Broderick et al. (2018).

In EEG preprocessing, we conducted a bandpass filtering of 0.5-40 Hz using a fourth-order zero-phase-shift Butterworth filter. The last seconds without audio input and the first five seconds of each trial were excluded to reduce filtering artifact. Noisy channels were then manually identified by examining data variance and abnormal values. On average, $0.33\pm0.94$ noisy channels were identified per trial in the Broderick dataset. Noisy channels were interpolated by the average of 3 adjacent channels. Independent component analysis was subsequently employed to eliminate the ocular and muscle artefacts, with 2-4 components removed for each subject. Ultimately, the data were re-referenced to the average of two mastoid channels. Noisetools (De Cheveigné & Arzounian, 2018) and Fieldtrip (Oostenveld et al., 2011) toolboxes were utilized in preprocessing.

*2.1.3  The FACED dataset of emotional video watching*

In the FACED dataset (Chen et al., 2023b), 32-channel EEG recordings were obtained from 123 subjects as they watched video clips. The EEG recordings had a sampling rate of 250 Hz. Participants viewed 28 video clips during the experiment, ranging in duration from 34 to 129 seconds (67 seconds on average). The video clips were chosen specifically to elicit neutral and eight distinct emotional states. The subjects were instructed to rate their emotional experiences in terms of valence, arousal, and the intensity of specific emotion categories at the end of each video clip.

For further details on the dataset, please refer to Chen at al. (2023b).

For EEG preprocessing, the data were first downsampled to 125 Hz, so that the sampling rates were roughly consistent between the two datasets. Then, the data was bandpass filtered between 0.5 to 40 Hz and segmented into trials. The subsequent denoising and re-referencing procedures followed the same steps and settings as those for the Broderick dataset. On average, 0.06±0.26 noisy channels were removed per trial. Thirty EEG channels (excluding two mastoid channels) were used for further analysis.

*2.2 The contrastive learning strategy*

Following the idea of ISC that there exist consistent, stimulus-evoked responses across subjects (Nastase et al., 2019), we design a contrastive learning model to filter out subject-specific signals and reveal maximally aligned responses. Specifically, we defined EEG responses of two subjects to an identical stimulus as a positive pair, whereas responses to different stimuli as negative pairs. Contrastive learning maximizes similarity between the positive pair in contrast to negative pairs. Consequently, the model transforms EEG signals into a space that maximizes similarity between responses of different subjects to the same naturalistic stimuli.

During the training phase of contrastive learning, data from two subjects were sampled together in a single mini-batch. For each trial, a continuous segment of duration $T$ was randomly sampled. The samples from the two subjects were precisely time-aligned, ensuring that the subjects were receiving identical stimulus content. Consequently, for each trial, two samples – one from each subject - were obtained, yielding $2N$ samples in a single mini-batch (where $N$ represents the number of trials). Each sample constituted a positive pair with its temporally aligned counterpart from the other subject. Samples from different time segments of two subjects formed $2(N-1)$ negative pairs.

To maintain diversity among the samples, a step size of *T*/2 was used for consecutive samples along the timeline. As a result, a total of $\frac{(T_{total}-T)}{T/2} + 1$ samples could be extracted from each trial. All possible positive sample pairs were enumerated within a single epoch.

The model was optimized by maximizing the similarity of representations between positive pairs relative to that between negative pairs. The InfoNCE loss (Oord et al., 2019) was adopted here:

$$l_i^A = -\log\left[\frac{exp(sim(z_i^A, z_i^B)/\tau)}{\sum_{j=1}^{N} \mathbb{1}_{[j \neq i]} exp\left(sim(z_i^A, z_j^A)/\tau\right) + \sum_{j=1}^{N} exp\left(sim(z_i^A, z_j^B)/\tau\right)}\right].$$

In the equation, $z_i^A$ ($i = 1,2,...,N$) is the representation of the sample from trial *i*, subject *A*. Similarly, $z_i^B$ ($i = 1,2,...,N$) is the representation of the sample from trial *i*, subject *B*. $sim(\cdot, \cdot)$ computed the cosine similarity between two representations as

$$sim(\boldsymbol{m}, \boldsymbol{n}) = \frac{\boldsymbol{m} \cdot \boldsymbol{n}}{\|\boldsymbol{m}\|\|\boldsymbol{n}\|}.$$

$\mathbb{1}_{[j \neq i]} \in \{0, 1\}$ is an indicator function, which is set to 1 iff $j \neq i$. $exp(\cdot)$ is the exponential function. $\tau$ is the temperature hyperparameter. Note that $z_i^A$ and $z_i^B$ came from the same trial *i* and formed a positive pair here. $z_i^A$ and $\{z_j^A, z_j^B \ (j \neq i)\}$ came from different trials and formed negative pairs. The model would maximize the similarity between $z_i^A$ and $z_i^B$ in contrast to that between $z_i^A$ and other samples. The total loss of a mini-batch was defined as

$$L = \frac{1}{N}\sum_{i=1}^{N} l_i^A + \frac{1}{N}\sum_{i=1}^{N} l_i^B,$$

which is the average loss after enumerating over each sample in the batch as the anchor sample.

*2.3 Model architecture and hyperparameters*

A convolution neural network was utilized to extract spatiotemporal EEG representations in CL-SSTER (Fig. 1, Fig. S1). Firstly, a spatial projection with a convolutional kernel size of $M \times 1$ was used to extract the latent spatial components, where $M$ is the number of EEG channels. $K_1$ kernels were utilized in the model, resulting in $K_1$ latent dimensions. After that, a temporal convolution with a kernel size of $1 \times P_1$ was employed to extract temporal information and decompose the signal into frequency bands. It functions similarly to a finite impulse response (FIR) filter with an order of $P_1$. Every temporal convolution filter was applied to all the latent spatial components, which means the parameters of the temporal convolution were shared across spatial components. A total of $K_2$ temporal convolution kernels were employed, leading to $K_2 \times K_1$ latent dimensions following the temporal convolution process. Lastly, the outputs underwent a ReLU activation function and were subjected to average pooling (with a size of $P_2$) over the temporal dimension before cosine similarities were calculated across the representations. The ReLU and average pooling operations approximate the instantaneous amplitude of the signal. Stratified normalization was employed to mitigate inter-subject discrepancies and facilitate the training process (Fdez et al., 2021). This technique involved concatenating the samples from a single subject within a mini-batch and normalizing each channel or dimension by subtracting the mean and dividing by the standard deviation. Stratified normalization was applied to the input EEG data and the representations after average pooling.

The hyperparameters utilized in our model are detailed in Table 1. The sample length ($T$) of the FACED dataset is smaller due to its relatively shorter trial durations. The sample length of over 5 seconds is generally adequate for capturing shared information of higher-level cognitive activities, such as emotion processing (Ouyang et al., 2022). To assess the influence of weight decay on model performance, we experimented with a range of values: 0, 0.01, 0.03, 0.1, 0.3, 1, 3, 10, 30, and 100. as demonstrated in Figs. S2, S3. Throughout other analyses in this paper, a weight decay value of 0.1 was empirically adopted to ensure satisfactory performance across all datasets and offered an appropriate level of regularization for the model parameters. We also tested the impact of the number of spatial and temporal filters, the average pooling size, and additional nonlinear layers on model performance (Fig. 7). The spatial and temporal filters were varied from 2 to 32 as geometric sequences with a common ratio of 2. The average pooling size was varied as 1, 5, 10, 15, 20 and 25. For the nonlinear layers, we added two convolutional layers with a kernel size of 3 and ReLU activations between them. Group convolution was utilized here, with one group for each kernel.

Table 1. Hyperparameter settings in CL-SSTER on the three datasets.

| Hyperparameters | Synthetic | Broderick (2019) | FACED |
|---|---|---|---|
| The number of spatial filters ($K_2$) | 16 | 16 | 16 |
| The number of temporal filters ($K_1$) | 16 | 16 | 16 |
| The temporal filter size ($P_1$) | 30 | 30 | 30 |
| The spatial filter size ($M$) | 128 | 128 | 30 |
| The average pooling size ($P_2$) | 15 | 15 | 15 |
| Weight decay | 0.1 | 0.1 | 0.1 |
| Training epochs | 30 | 50 | 50 |
| Learning rate | 0.0007 | 0.0007 | 0.0007 |
| Temperature in loss function ($\tau$) | 0.07 | 0.07 | 0.07 |
| Sample length ($T$) | 30 seconds | 30 seconds | 5 seconds |

To assess the model's performance, we conducted leave-one-subject-out cross-validation for the synthetic dataset and the Broderick dataset. For the FACED dataset, we used ten-fold cross-validation due to its large sample size. The cosine similarity across different samples was visualized in Fig. S4 to demonstrate the effects of contrastive learning on sample similarity.

*2.4 ISC computation and model interpretation*

*2.4.1 Postprocessing of model outputs*

The model can achieve a higher ISC with an adequate number of dimensions in our experiment (Fig. 7); however, this also results in more highly correlated dimensions. To extract succinct and uncorrelated components from the model outputs, clustering was performed on the dimensions of model's outputs based on their correlation matrix. We calculated the Pearson correlation coefficients between each pair of the latent dimensions within a trial, then averaged these correlation matrices across trials and subjects. The affinity matrix for the dimensions was defined as one minus the absolute values of the average correlation matrix. Agglomerative hierarchical clustering was employed to derive a dendrogram of the dimensions based on the affinity matrix. We used sklearn.cluster.AgglomerativeClustering function with the average linkage criterion implemented in Python. Dimensions with an absolute mean correlation below 0.1 were categorized into separate clusters, as a correlation below 0.1 is generally considered negligible in the context of correlation analysis (Schober et al., 2018). The clustering analysis was conducted based on the CL-SSTER model trained on all data in each dataset. The dimension with the highest inter-subject correlation (ISC) in each cluster was selected as the representative dimensions for further analysis.

*2.4.2    The computation of ISC*

ISC for each dimension or component was calculated to quantify its inter-subject consistency. For this, the Pearson correlation coefficients between dimensions within the same trial from two subjects were computed, then these coefficients were averaged over all trials and all subject pairs. In the computation of ISC for the training data, all pairs of subjects in training data were included. For validation data, a validation subject's ISC was determined by averaging their ISC with that of all other subjects in the dataset, for a given dimension. In the paper, the average and standard error of ISC across all validation subjects were reported.

To determine the ISC of the representative dimensions on the validation sets, we identified the corresponding dimensions in each fold of cross-validation. Specifically, the representative dimensions were identified from clustering based on the model trained on all data. For each fold of the cross-validation, we selected the output dimensions that had the highest correlation with these representative dimensions. The ISC for these correlated dimensions on the validation sets was then computed. The average ISC of these dimensions across all folds of the cross-validation was reported as the ISC of the representative dimensions in the paper.

We conducted permutation tests to determine whether a component was significantly correlated across subjects. Random circular shifts were employed to permute the data, whereby the entire time series of EEG signals from one subject were shifted by a random index $o$ (Dmochowski et al., 2012; Parra et al., 2019). Formally, assuming $X \in \mathbb{R}^{M \times T}$ represents the EEG signals from one subject, the values of $X_{:,mod(i+o,T+1)}$ were assigned to $X_{:,i}$, where $mod$ is the modular operation. The data from each subject were shifted by different indexes $o$. The random circular shift disrupted the temporal alignment across subjects but preserved the basic spatial and spectral properties of the original data. The permuted data were then inputted to the CL-SSTER method to obtain the null distribution of ISC values.

*2.4.3   Analysis of the convolutional kernels*

The temporal convolution and spatial projection parameters in the model were analyzed to identify the temporo-frequency and spatial characteristics of the learned representations. The temporal convolution was expected to filter the signals into different frequency bands. To characterize the frequency response of temporal convolutional kernels, a fast Fourier transformation was applied to the kernel parameters and the modulus in the frequency domain was taken to represent the spectrum magnitude. The frequency ranges with high spectrum magnitudes indicate the passbands of the temporal convolutional kernel (see the bottom row of Fig. 3c for an example).

The spatial projection identified latent components by linear combination of signals from different channels. According to Haufe et al. (2014), parameters of spatial projections (or spatial filters) should be transformed to spatial activation to represent the interpretable spatial distributions of the latent components. The spatial activations were calculated by $A_k = \overline{\Sigma_X} W_k$ $(k = 1, 2, \ldots, K_1)$, where $W_k$ represented the spatial filters and $\overline{\Sigma_X}$ denoted the average covariance of EEG signals across all subjects (Haufe et al., 2014). The spatial activations $A_k$ represented the optimal parameter for recovering EEG signals from the latent space, commonly referred to as parameters in a forward model (See the right columns of Fig. 5b for examples of spatial filters and spatial activations). Here, we calculated $A_k$ for each latent dimension separately without considering their interactions.

*2.5 The relationship between EEG representations and stimuli characteristics*

The learned EEG representations are expected to reflect the underlying neural processing of naturalistic stimuli that are common across subjects. Therefore, we aimed to identify the relationship between EEG representations and stimulus characteristics. For features that exhibit continuous variation over time, including speech envelope, semantic dissimilarity, and video brightness, we implemented a temporal response function analysis to quantify their predictive power for EEG representations following previous studies (Broderick et al., 2018; Li et al., 2022; Li et al., 2023). For stimulus characteristics annotated on a trial-by-trial basis, including emotional valence and arousal, we calculated their correlation with representative EEG representations. It should be noted that the complexity of information in naturalistic stimuli makes it challenging to encompass all possible attributes. To evaluate the model's capacity for extracting meaningful representations, we selected a range of representative features. These span from the low-level features such as speech envelope and video brightness to more complex, high-level features like semantic dissimilarity and emotional content.

### 2.5.1 *The temporal response function analysis for speech envelope, semantic dissimilarity and video brightness*

The forward TRF models were constructed to predict the brain responses from the input stimulus features through linear convolution. The model weights describe the relationship between the stimulus features and the neural responses for a range of time lags. The prediction accuracy was quantified by the Pearson correlation coefficient (R) between the predicted responses and the actual responses, which indicates how well the stimulus information was encoded. The TRF analysis was implemented by the mTRF Matlab Toolbox (Crosse et al., 2016).

For the Broderick dataset, we extracted amplitude envelope and semantic dissimilarity features of the speech. The amplitude envelope was computed by performing Hilbert transformation on the stimulus and taking the absolute values. The semantic dissimilarity calculation followed the procedures in Broderick et al. (2018). We first transformed each word into a 300-dimension vector using the word2vec function in Matlab. Then, for each word, we calculated the Pearson correlation coefficient between its vector and the average vector of all preceding words in the sentence. Subtracting the correlation coefficient from one yielded the semantic dissimilarity value. For the FACED dataset, we chose brightness as the stimulus feature, which was averaged for each frame of the video. All the stimulus features underwent resampling to match the sampling rate as the EEG data.

TRF models were constructed to predict the original EEG responses and the CL-SSTER representations from the speech features, respectively. The inputs of TRF models were data from complete trials without epoching to avoid edge artefacts. We then compared the correlation coefficients (R values of prediction) in these two TRF models, $R_{original}$ and $R_{CL-SSTER}$. The TRF analyses were conducted on two levels: the individual-level and the group-level. For the individual-level analysis, the TRF models were trained subject-wise and the prediction Rs were averaged across subjects. For the group-level analysis, the original EEG responses or the CL-SSTER representations were averaged across subjects. A single holistic TRF model was trained to predict either the group-averaged original responses or the group-averaged representations.

*2.5.2   The correlation analysis for emotional ratings*

We conducted correlation analysis between EEG representations and emotional ratings to see whether the learned inter-subject shared representations can reflect subjects' emotional experience of valence and arousal better than the original signals. Individual-level and group-level analysis were both conducted. At the individual-level, the subject-wise EEG features or learned representations were correlated with the individual ratings of valence or arousal, then R values were averaged across all subjects. At the group-level, the EEG features or CL-SSTER representations were averaged across subjects. The ratings also underwent a group averaging process. A correlation analysis was then conducted between the group-averaged features/representations and the ratings. For the learned representations, power spectral density (PSD) features of each trial were extracted from the representative components before ReLU activation and average pooling. For comparison, PSD features of the original signals were extracted from each EEG channel and each frequency band (broadband, delta, theta, alpha, beta and gamma). In each frequency band, the highest correlation coefficient among all channels was reported.

*2.6 Comparison methods*

We compared our method with the correlated component analysis (CorrCA) and ISC analysis of the original signals. CorrCA represented a state-of-the-art approach widely recognized in the realm of intersubject correlation analysis (Dmochowski et al., 2012). CorrCA utilized a linear projection of EEG channels to maximize inter-subject correlation within the latent space. Traditionally, the CorrCA methodology was applied to broadband EEG signals. In our investigation, we filtered the EEG data into five frequency bands (delta: 0.5-4 Hz, theta: 4-7 Hz, alpha: 7-12 Hz, beta: 12-30 Hz, gamma: 30-40 Hz), thus facilitating a more equitable comparison with our proposed method. Cross-validation using the same partitioning of training and validation data as the CL-SSTER method was performed for CorrCA. The average of the maximal ISCs across all cross-validation folds in each frequency band (broadband and five decomposed bands) was reported. Permutation tests the same as in *2.4.2* were applied to CorrCA to determine whether the ISC of each component is significantly larger than the permuted data. The spatial activation patterns of components with significant ISCs were visualized in Figs. S6, S8. For ISC analysis of the original signals, we also decomposed the EEG data into the five frequency bands and reported the maximal ISCs of these bands as well as the broadband.

# 3 Results

*3.1 Performance validation on the synthetic dataset*

On the synthetic dataset, we evaluated the performance of the CL-SSTER method under three criteria. Firstly, the CL-SSTER method identified the shared spatial activations more accurately than CorrCA did (Fig. 3b). The Pearson correlation coefficients (Rs) between the ground-truth shared spatial patterns and the spatial activations of the CL-SSTER method reached 0.984, 0.985, 0.962 and 0.995 for the four frequency ranges, higher than those of the CorrCA method (0.920, 0.834, 0.730, 0.955). Secondly, the temporal filters in the trained CL-SSTER model had similar frequency responses as the generated narrowband oscillations (Fig. 3c). The peak frequency responses of the temporal filters (the bottom row in Fig. 3c) exhibited a high degree of correspondence to the prominent peaks in the power spectral distribution of the inter-subject shared response (Fig. 3a). Thirdly, the latent representations extracted by CL-SSTER achieved maximal ISCs of $0.341\pm0.001$, $0.328\pm0.000$, $0.505\pm0.003$, $0.521\pm0.010$ for the four frequency ranges, which were significantly higher than those of the original signal ($0.008 \pm 0.001$, $0.010 \pm 0.001$, $0.011 \pm 0.001$, and $0.011\pm0.001$), and those of the CorrCA method ($0.049\pm0.002$, $0.039\pm0.003$, $0.031\pm0.002$, and $0.065\pm0.002$, $p < 0.001$ for all paired-sample t-tests. See Fig. 3d). Further, we tested the reliability of the CL-SSTER method by conducting 10 separate executions with different random seeds. The spatial activations and frequency responses of the filters showed a high degree of consistency in all ten executions (Fig. 4), as reflected by the Pearson correlation coefficients exceeding 0.84 for spatial activations and 0.90 for frequency responses, confirming the method's reliability. In summary, our results on the synthetic dataset prove that the CL-SSTER method can effectively and reliably detect spatial and temporal patterns of the shared responses and draw out the expected inter-subject shared component from noisy data.

To validate the effectiveness of contrastive learning in minimizing the difference between positive sample pairs, we computed the cosine similarity between EEG representations across different trials and subjects. The similarity of the same trials across subjects (auxiliary diagonals in Fig. S4) became apparently higher than those of different trials after training, indicating the method can effectively maximize the similarity between positive sample pairs.

*3.2 Performance validation on the natural speech comprehension EEG dataset*

On the Broderick dataset of naturalistic speech, we identified two representative components through clustering on the output dimensions, one representing dimensions with lower frequency response and the other with higher frequency response (Fig. 5b, Fig. S5). These components exhibited ISCs of 0.050±0.003 and 0.014±0.002 respectively (Fig. 5a), both significantly higher than ISCs of the randomly shuffled data in the permutation tests ($ps < 0.05$). The maximal ISC of the components identified by CL-SSTER was significantly higher than that of the original signals (delta band, maximal ISC: 0.022±0.000, $t(18) = 8.69$, $p < 0.001$, paired-sample t-test), as well as that of the components identified by CorrCA method (delta band, maximal ISC: 0.017±0.002, $t(18) = 11.20$, $p < 0.001$). Note that we decomposed the data into different frequency bands for original ISC and CorrCA analysis for a relatively fair comparison with the CL-SSTER method. Broadband data, which was generally used in classical analysis, produces ISC even lower than that in specific bands. Therefore, CL-SSTER can extract consistent response patterns across subjects better than the CorrCA method.

The temporal and spatial filters learned by CL-SSTER were visualized in Fig. 5b. The two components had frequency responses of < 2 Hz and 6-11 Hz, respectively. The first component overlapped partially with the delta band. The second component overlapped with part of the theta and alpha bands. The original EEG signals in these matching frequency bands also had higher ISCs than in other bands, which indicates that the model can basically identify the frequency ranges of the inter-subject shared activities. Furthermore, the CL-SSTER method can single out a flexible frequency range rather than predefined bands.

For the spatial filters, channels in frontotemporal and frontocentral regions have the highest weights (Fig. 5b), which are consistent with previous findings of regions correlated with speech information (Défossez et al., 2023; Di Liberto et al., 2015). The spatial activation of the first component displays inverse activations in the parietooccipital region and the frontotemporal region, which resembles the first component identified by the CorrCA method in broadband and delta band (Fig. S6), but the negative activation is more posterior in CL-SSTER. Unlike CorrCA, CL-SSTER jointly optimized temporal and spatial filters, thus producing a slightly different activation pattern, which possibly gave rise to the higher ISC achieved.

Utilizing the learned inter-subject shared components, we conducted temporal response function (TRF, Crosse et al. 2021) analyses to identify their relationship with amplitude envelope and semantic dissimilarity of the speech. For both group-level and individual-level analyses, component 1 derived from CL-SSTER demonstrated a more effective encoding of speech features, as evidenced by higher R values compared to the original EEG signals across all frequency bands (Fig. 6a, b). In the group-level TRF modelling, the R value associated with predicting component 1 in CL-SSTER from amplitude envelope reached 0.405, in contrast to only 0.120 when predicting the original EEG signals. Likewise, for predictions based on semantic dissimilarity, the R values were 0.192 for component 1 in CL-SSTER and 0.062 in the case of the original EEG signals. In the individual-level analysis, component 1 in CL-SSTER (envelope: $0.130 \pm 0.011$, semantic dissimilarity: $0.061 \pm 0.005$) also obtained significantly higher R values than the original EEG signals (delta band, envelope: $0.065 \pm 0.008$, semantic dissimilarity: $0.026 \pm 0.004$) when predicted from both amplitude envelope ($t(18) = 8.79$, $p < 0.001$, paired-sample t-test) and semantic dissimilarity ($t(18) = 5.21$, $p < 0.001$). Component 1 consistently demonstrated the best performance in the TRF model when predicted from both amplitude envelope and semantic dissimilarity. This suggests that it may serve as a comprehensive representation for inter-subject shared speech processing.

Further, we explored the effects of varying hyperparameters and model architectures on the results. Firstly, we changed the number of spatial and temporal filters in the model to observe their effects on validation ISC values. The results indicated a sufficient number of spatial and temporal filters are necessary to achieve higher ISC (Fig. 7a), despite we clustered on the output dimensions to get succinct representational dimensions. Generally, using more than 4 temporal filters and more than 2 spatial filters yielded satisfactory performance, suggesting that the model is not highly sensitive to the exact number of filters. The selected hyperparameters (16 spatial filters and 16 temporal filters) yielded ISCs of with no significant difference from other settings within this range ($p > 0.05$, paired sample t-test, FDR corrected), except that two settings (16 temporal filters and 4 spatial filters, and 32 temporal filters and 8 spatial filters) were significantly lower than the selected hyperparameters. Secondly, we varied the window size of average pooling in the model and found that a pooling size of 15 yielded the best performance (Fig. 7b), likely because a window of around 1/10 second is appropriate for approximating the amplitude. It should be noted that the performance is also not highly sensitive to the average pooling size. Among the evaluated settings, only the performance with an average pooling size of 5 was significantly lower than the that of 15 ($p = 0.010$, paired-sample t-test, FDR corrected). Thirdly, we added two convolutional layers to the model to evaluate their impact on the performance (see Section 2.3 for details). This modification yielded a lower ISC value ($0.035 \pm 0.002$ compared to $0.050 \pm 0.003$ without these layers, $p < 0.001$, paired-sample t-test Fig. 7c), indicating simply adding more nonlinear layers cannot improve the model's performance.

*3.3 Performance validation on the emotional video watching EEG dataset*

For the FACED dataset with emotional videos as stimuli, we identified six components from the clustering analysis of the model's outputs. Their representative filters have frequency response of 0-2 Hz, 0-2 Hz, 6-10 Hz, 2-6 Hz, 10-19 Hz and 2-6 Hz respectively (Fig. 8b, Fig. S7). ISCs of these filters are all significantly higher than those of the randomly shuffled data in the permutation tests ($p$s < 0.05). The component with maximal ISC (0.075±0.002) outperformed the ISCs of the original signals (delta band, maximal ISC: 0.047±0.001, $t(122) = 17.66$, $p < 0.001$, paired-sample $t$-test) and is comparable to those of the CorrCA method (delta band, maximal ISC: 0.074±0.002, $t(122) = 1.20$, $p = 0.230$) (Fig. 8a). The maximal ISC of the CL-SSTER method was much higher than that of the original signals (0.028±0.001, $t(122) = 28.10$, $p < 0.001$, paired-sample $t$-test) and of CorrCA (0.043±0.001, $t(122) = 22.75$, $p < 0.001$) in the broadband.

CL-SSTER identified components across different frequency ranges, with the component in low delta band (0-2 Hz) achieving the highest ISC. Moreover, the learned components generally produced higher ISCs than their opponents / the corresponding components of CorrCA method with similar frequency ranges. For example, component 3 with the frequency range of 6-10 Hz has a much higher ISC (0.038±0.001) than that of CorrCA method in theta band (0.017±0.001, $t(122) = 27.24$, $p < 0.001$, paired-sample t-test) or alpha band (0.004±0.0002, $t(122) = 36.67$, $p < 0.001$). It again indicated the temporal and spatial filters flexibly learned by CL-SSTER better characterized inter-subject shared activities than the CorrCA method.

The spatial filters learned by CL-SSTER showed various activations spanned from frontal to occipital regions (Fig. 8b). Component 1, which produces the maximal ISC, showed an anterior-posterior activation pattern. Component 2 displayed a more anterior activation in the middle fronto-central region. The remaining four components mainly displayed activations in occipital and parietal regions. The spatial patterns of the first two components resembled the counterparts of CorrCA in the broadband and the delta band (Fig. S8).

Then we identified the relationship between EEG representations and stimuli features, i.e., video brightness and emotional valence and arousal features. In comparison with the original EEG signals, the CL-SSTER representations demonstrated more effective encoding of all video features both at the individual and the group level (Fig. 9). For the group-level analysis (Fig. 9, left column), the best R value for predicting CL-SSTER representations from brightness was 0.140 (component 3), higher than the best R value for predicting the original signals (0.082, broadband). The best absolute R values of correlation with arousal or valence were also higher for CL-SSTER representations (arousal: 0.493 for component 2, valence: 0.440 for component 3) than the original signals (arousal: 0.321 for the beta band, valence: 0.376 for the theta band). For the individual-level analysis (Fig. 9, right column), CL-SSTER representations outperformed the original signals in correlations with brightness (CL-SSTER: $0.023 \pm 0.002$ for component 3, original signals: $0.006 \pm 0.001$ for the broadband, $t(122) = 9.63$, $p < 0.001$, paired-sample t-test), arousal (CL-SSTER: $0.093 \pm 0.021$ for component 4, original signals: $0.080 \pm 0.020$ for the delta band, $t(122) = 4.00$, $p < 0.001$) and valence (CL-SSTER: $0.176 \pm 0.019$ for component 3, original signals: $0.170 \pm 0.017$ for the delta band, $t(122) = 7.54$, $p < 0.001$). These results suggested that the components learned by CL-SSTER can be more effectively predicted by the stimulus information than the original signals.

Fig. 9 illustrated that different components learned by CL-SSTER captured distinct neural processes during video watching. Components 3 and 5, with occipital activations, were most effectively predicted by brightness feature, suggesting its relation to low-level visual processing.

Components 2 and 4 demonstrated the highest correlations with arousal, while components 3 and 6 exhibited the most pronounced correlations with valence, which implied that they encapsulated the emotional experiences of the participants.

## 4 Discussion

The present study proposed a contrastive learning framework for identifying inter-subject shared spatiotemporal representations. The framework was systematically validated on three datasets, including a synthetic dataset and two real-world datasets with naturalistic stimuli. The method consistently outperformed the competing CorrCA method by identifying components with higher inter-subject correlations. The learned representations exhibited specific frequency responses and spatial activities that align with existing literature. The learned representations were capable of tracking the real-time, continuous variations of stimulus properties, such as speech amplitude envelope, semantic dissimilarity, and video brightness, and also exhibited correlations with the emotional features of the stimulus. The contrastive learning framework emerged as a powerful and extensible tool for inter-subject correlation studies, unveiling the potential for mining subtle but meaningful inter-subject shared components from EEG signals.

Through both the synthetic and the real-world datasets, we demonstrated that the proposed method can substantially raise ISC values. Currently, the ISC values reported in literature are relatively modest, with the ISCs achieved by CorrCA less than 0.05 (Ki et al., 2016; Parra et al., 2019; Rosenkranz et al., 2021). We showed on the two real-world datasets that CL-SSTER can raise the ISCs by an average of 81.2% and 37.4% in comparison to ISC of the sensor space and of the CorrCA method, respectively. The higher ISC values indicated the method can identify neural representations that are more consistent across subjects. It could provide a more sensitive measure for the level of inter-subject correlations in inter-brain studies.

The results also demonstrated that CL-SSTER can automatically extract the critical temporal and spatial patterns in naturalistic settings. Without prior filtering EEG signals to fixed bands, CL-SSTER can detect frequency ranges with a flexible boundary, potentially shedding light on previously overlooked information. For the synthetic data, we generated EEG oscillations in physiologically possible yet untraditional bands ranging from lower to higher frequencies, and showed that CL-SSTER can successfully single out these bands with a high precision (Fig. 3c). On the real-world datasets, the original EEG signals and the CorrCA method generally produced poor ISCs in higher frequency bands. In contrast, ISCs achieved by CL-SSTER in higher frequency bands were nontrivial and were generally higher than competing methods in similar bands (Fig. 8). Besides identifying relevant frequency bands, CL-SSTER also revealed spatial patterns that fit into existing literature. The first component obtained from the Broderick dataset is similar to one of the significant components found in previous studies incorporating speech or speech-video stimuli (Cohen et al., 2017; Cohen & Parra, 2016; Rosenkranz et al., 2021), with inverse activations in parieto-occipital and bilateral frontotemporal regions. On the FACED dataset, the first component with anterior-posterior activation resembles the microstate C in literature (Michel & Koenig, 2018), which is one of the topological maps that explained the largest variance of EEG data (Gui et al., 2020; Liu et al., 2023; Zanesco et al., 2020). This component was also identified in previous ISC studies on moving watching (Dmochowski et al., 2014; Dmochowski et al., 2012; Petroni et al., 2018; Poulsen et al., 2017). Furthermore, as CL-SSTER jointly optimized temporal and spatial filters, it can reveal specific combinations of temporal and spatial patterns with significance, and further validate or extend previous findings. For example, previous research demonstrated that the component with negative bilateral frontotemporal activations was uniquely evoked by auditory stimuli, thus possibly related to auditory processing (Cohen & Parra, 2016). Our finding further validates the point by pinning down its frequency response at 0.5-2 Hz, corresponding to a range where neural tracking of continuous speech emerges (Ding & Simon, 2012, 2014; Horton et al., 2013; Keitel et al.; Zhang et al., 2023). This component may reflect neural tracking of the speech stimulus consistent across subjects. Taken together, these results illustrated that the proposed method can flexibly and effectively cope with the diverse inter-subject neural patterns under naturalistic scenarios.

It should be noted that ISCs of the latent space on the synthetic dataset were much higher than that on the two real-world datasets, possibly due to that the shared responses were not exactly the same across subjects in reality. The significant advancement of CL-SSTER over CorrCA method on the synthetic dataset indicated its ability of extracting the shared responses whenever they exist, even if ISC in the sensor space was as low as approximately 0.01. Comparing with the Broderick dataset, ISC of the sensor space was much higher on the FACED dataset, indicating that videos evoked more consistent responses across subjects than speech as they contained multimodal visual and auditory information. On the FACED dataset, CorrCA method produced higher ISCs than those in the sensor space, but on the Broderick dataset, it even fell behind. This could suggest a potential limitation of CorrCA in identifying more subtle inter-subject correlated components, as these components might have flexible temporal or frequency patterns. In contrast, CL-SSTER yielded much higher ISC on the Broderick dataset, than both CorrCA and the original EEG signals, manifesting its strength in more challenging scenarios.

Last but not least, the CL-SSTER method has been made open-source as a Python package, featuring a user-friendly interface for easy invocation. Detailed documentation is provided for the input data format and hyperparameters. Furthermore, post-hoc analysis and visualization are exemplified in Jupyter notebook files. This toolbox can be utilized in a broad range of inter-brain studies, enabling the identification of components that traditional methods might overlook.

In the future, researchers can delve into the group differences in brain synchronization patterns. By applying this approach, researchers can quantitatively assess and compare how different populations, perhaps varying in cognitive abilities or neurological conditions, exhibit unique synchronization characteristics (Byrge et al., 2015; Finn et al., 2018; Gao et al., 2020; Guo et al., 2015; Hasson et al., 2009; Iotzov et al., 2017; Jangraw et al., 2023; Yang et al., 2020). This could contribute to the development of tailored interventions or therapies based on specific synchronization patterns observed in different groups. Besides, the present study exemplifies the application of the CL-SSTER method to EEG data, future research could broaden this approach to encompass other modalities such as fMRI and functional Near-Infrared Spectroscopy (fNIRS). This expansion would position the method as a versatile instrument for cognitive neuroscience research, particularly in naturalistic settings.

**Data availability.** The data for Broderick et al. (2019) are available under a CC0 1.0 Universal (CC0 1.0) Public Domain Dedication license. The audio files for Broderick et al. (2019) were provided by the authors. The data for the FACED dataset are available under a Creative Commons Attribution 4.0 International License. The synthetic data and the preprocessed data of Broderick dataset and FACED dataset used in the paper are available at

https://www.synapse.org/#!Synapse:syn53641899/files/.

**Code availability.** The source code for implementation of the contrastive learning model and analysis of the representations is available at https://github.com/kekehia123/cl_sster/tree/main.

Acknowledgements

This work was supported by the National Natural Science Foundation of China (T2341003, 61977041), the National Key R&D Program of China (2021YFF1200804), National Natural Science Foundation of China (62001205), Shenzhen Science and Technology Innovation Committee (2022410129, KCXFZ2020122117340001, RCBS20231211090748082, KJZD20230923115221044), the Teaching Reform Project of the Instruction Committee of Psychology in Higher Education by the Ministry of Education of China (20222008), and the Education Innovation Grants of Tsinghua University (DX05_02).

Author contributions

D.Z. and X.S. conceptualized the project. D.Z. and Q.L. jointly supervised the project. X.S. and L.T. implemented the contrastive learning method and conducted the majority of the experiments. X.C. participated in the implementation of the temporal response function. S.S. contributed to the proposal of the contrastive learning strategy. X.S., L.T., D.Z., and Q.L. contributed to manuscript preparation. All authors contributed to manuscript revisions.

Competing interests

The authors declare no competing interests.

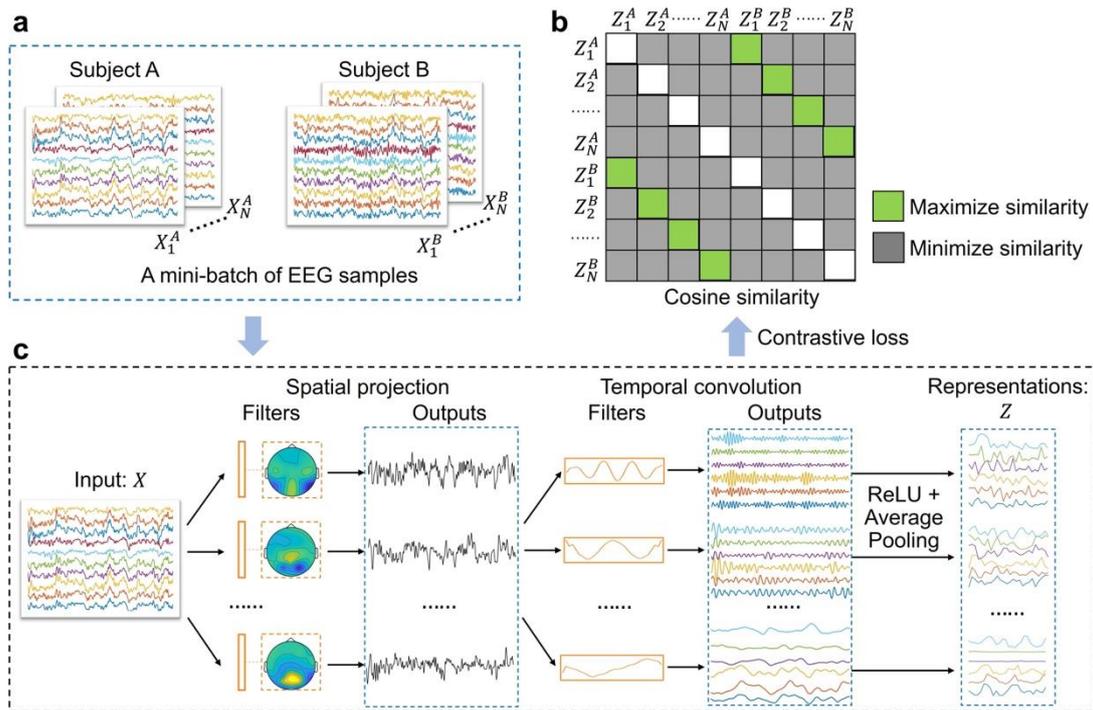

**Fig. 1 | Illustration of the CL-SSTER (Contrastive Learning for Shared SpatioTemporal EEG Representations) framework. a** A mini-batch of EEG samples were extracted from two subjects. **b** The samples that corresponded to the same stimuli received by the two subjects were regarded as a positive pair. The similarity of their latent representations ($Z_i^A$ and $Z_i^B$, $i = 1,2,...,N$) were maximized. **c** The model contained a spatial projection layer and a temporal convolution layer to extract the inter-subject shared spatial and temporal patterns, respectively.

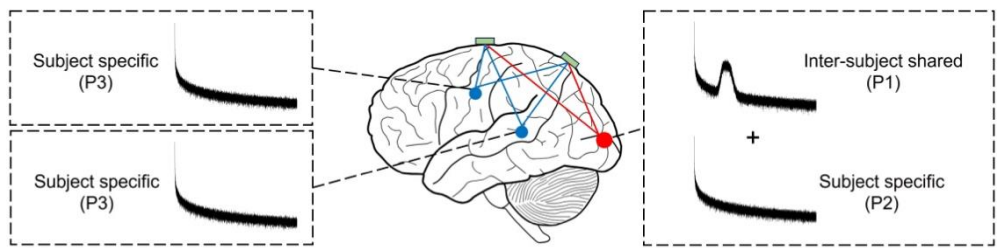

**Fig. 2 | Composition of the synthetic data.** The synthetic data consist of an inter-subject shared component (P1), a subject-specific component from the same source (P2), and subject-specific components from other sources (P3).

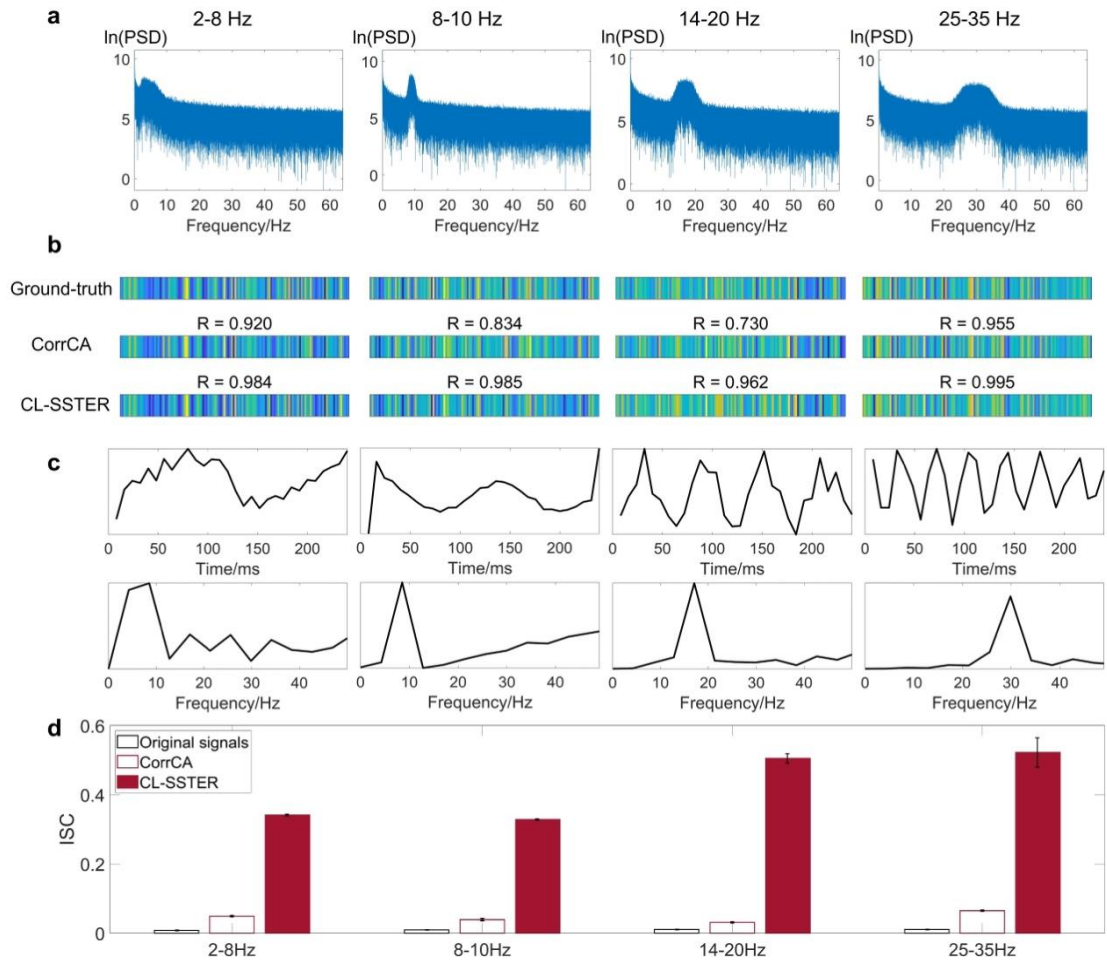

**Fig. 3 | Model validation on the synthetic dataset. a** Frequency responses of the inter-subject shared component. **b** Model performance of identifying ground truth spatial activations. CL-SSTER achieved a higher accuracy than CorrCA. **c** Temporal filters (on the top row) and their frequency responses (on the bottom row) in the CL-SSTER model. Frequency responses of the filters matched frequency peaks in the synthetic signals. **d** ISCs achieved by different methods. Maximal ISCs of the CL-SSTER method were significantly higher than that of the original signal and the CorrCA method.

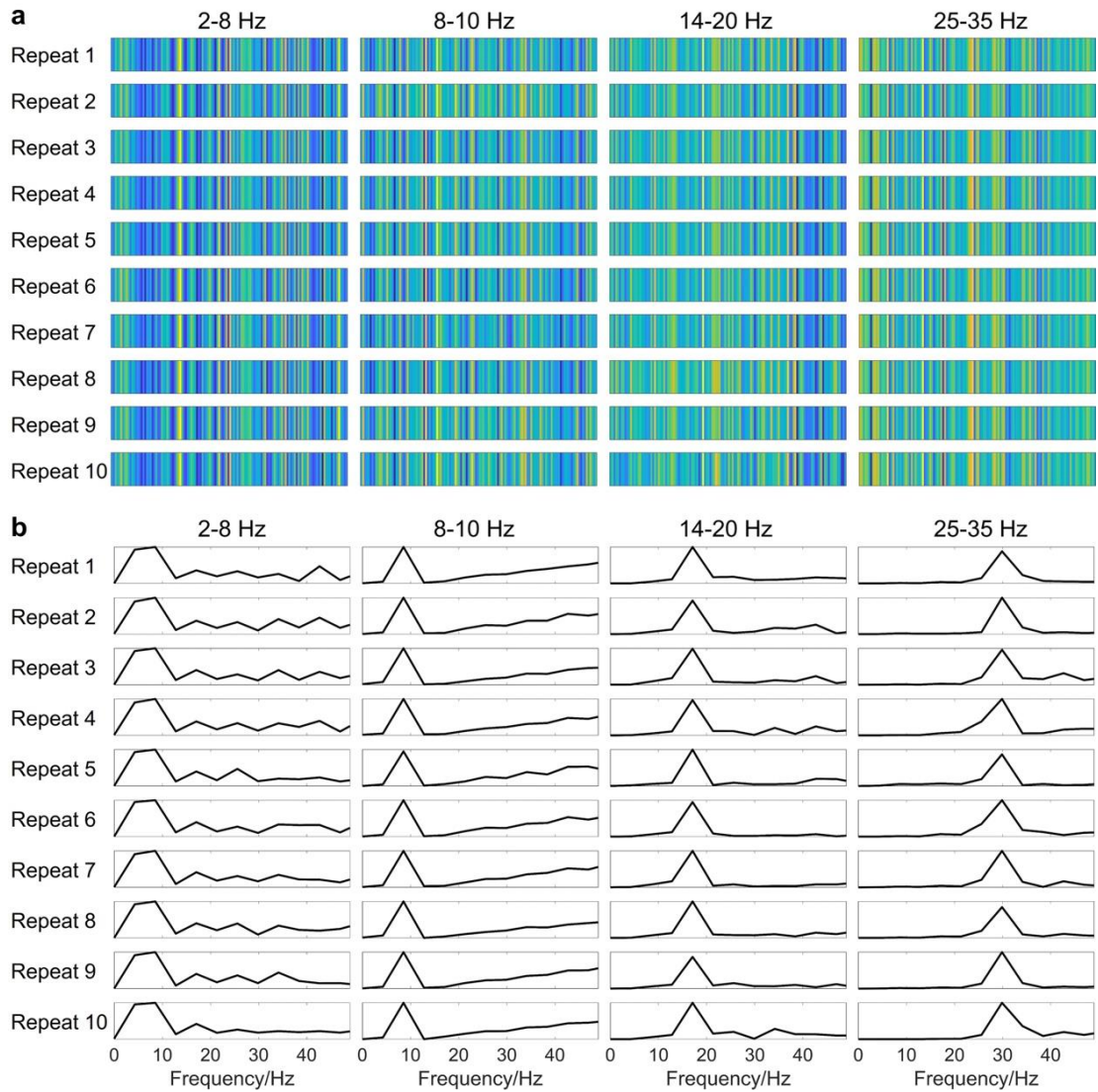

**Fig. 4 | Reliability analysis of the CL-SSTER method.** Spatial activations (a) and frequency responses (b) of 10 repeated executions of the model, each using a different random initialization seed, were presented. The component with the largest ISC was visualized for each execution. The Pearson correlation coefficient of the spatial activations between any two executions were above 0.84. The Pearson correlation coefficient of the frequency responses between any two executions were above 0.90.

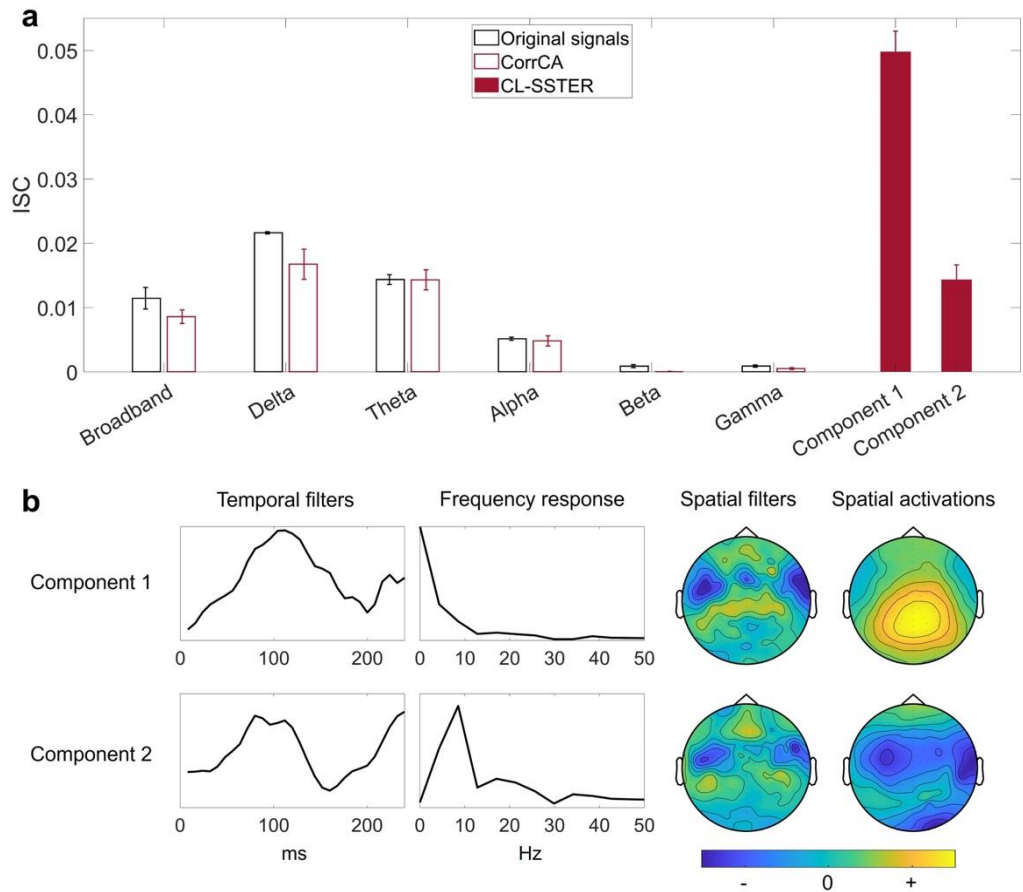

**Fig. 5 | ISCs of inter-subject shared components and their spatiotemporal characteristics on the Broderick dataset. a** ISCs comparison of different methods. The maximal ISCs of the EEG signals and of the CorrCA components in each frequency band were shown as the left hollow bars. The ISCs of the 2 components identified by CL-SSTER were shown as the right filled bars. **b** Visualization of representative temporal and spatial filters learned by CL-SSTER on the Broderick dataset.

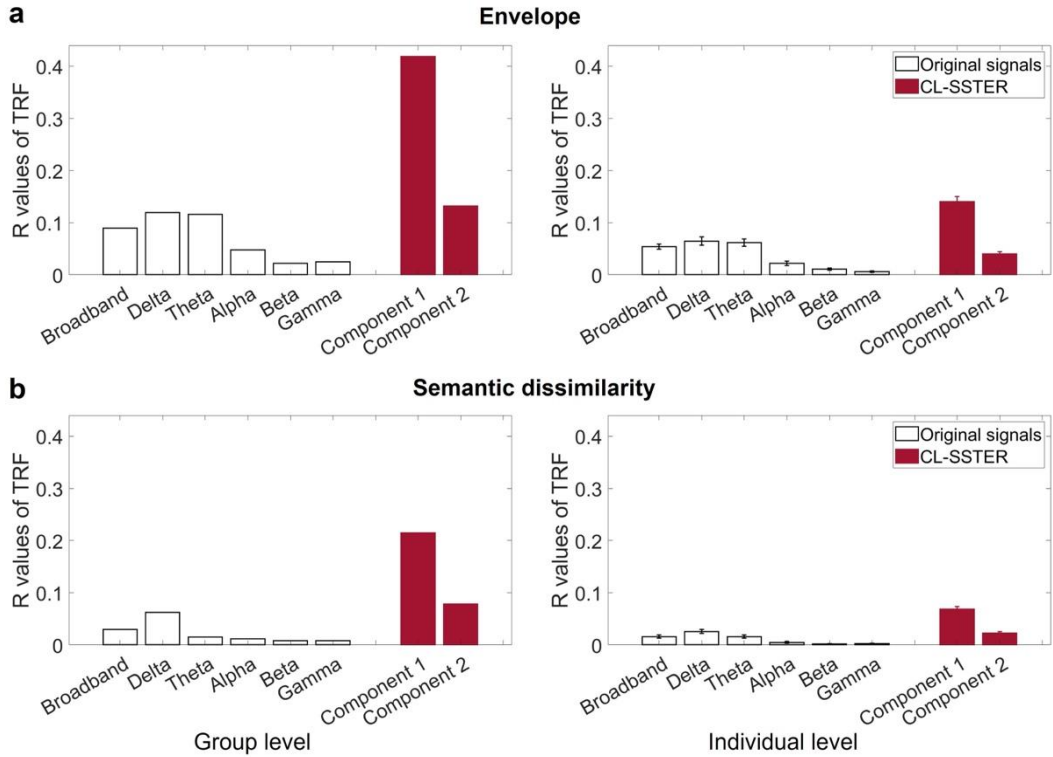

**Fig. 6 | Temporal response function (TRF) analysis on the Broderick dataset. a** R values of TRF models predicting EEG signals or learned representations of CL-SSTER from amplitude envelope. **b** R values of TRF models predicting EEG signals or learned representations of CL-SSTER from semantic dissimilarity. Left column: group level. Right column: individual level.

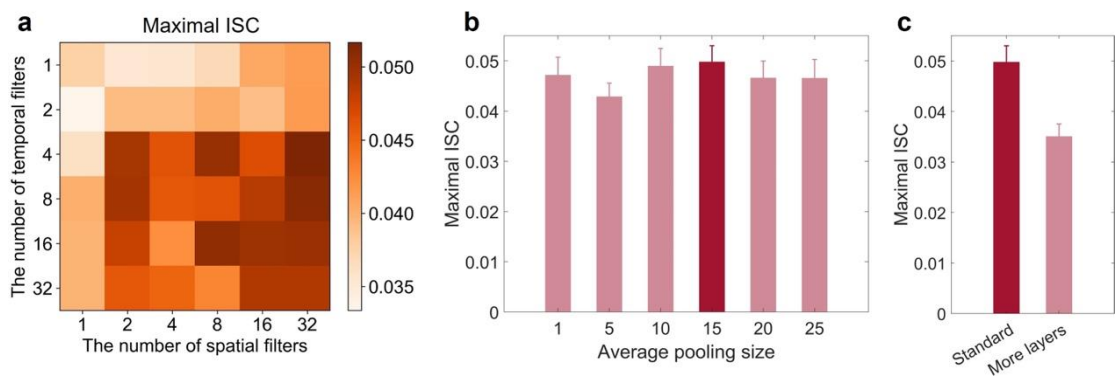

**Fig. 7 | The effects of varying hyperparameters and model architectures on maximal inter-subject correlation (ISC). a** Maximal ISCs with different numbers of spatial and temporal filters. **b** Maximal ISCs with different average pooling sizes. **c** Maximal ISCs of the standard model and the model with more convolutional layers.

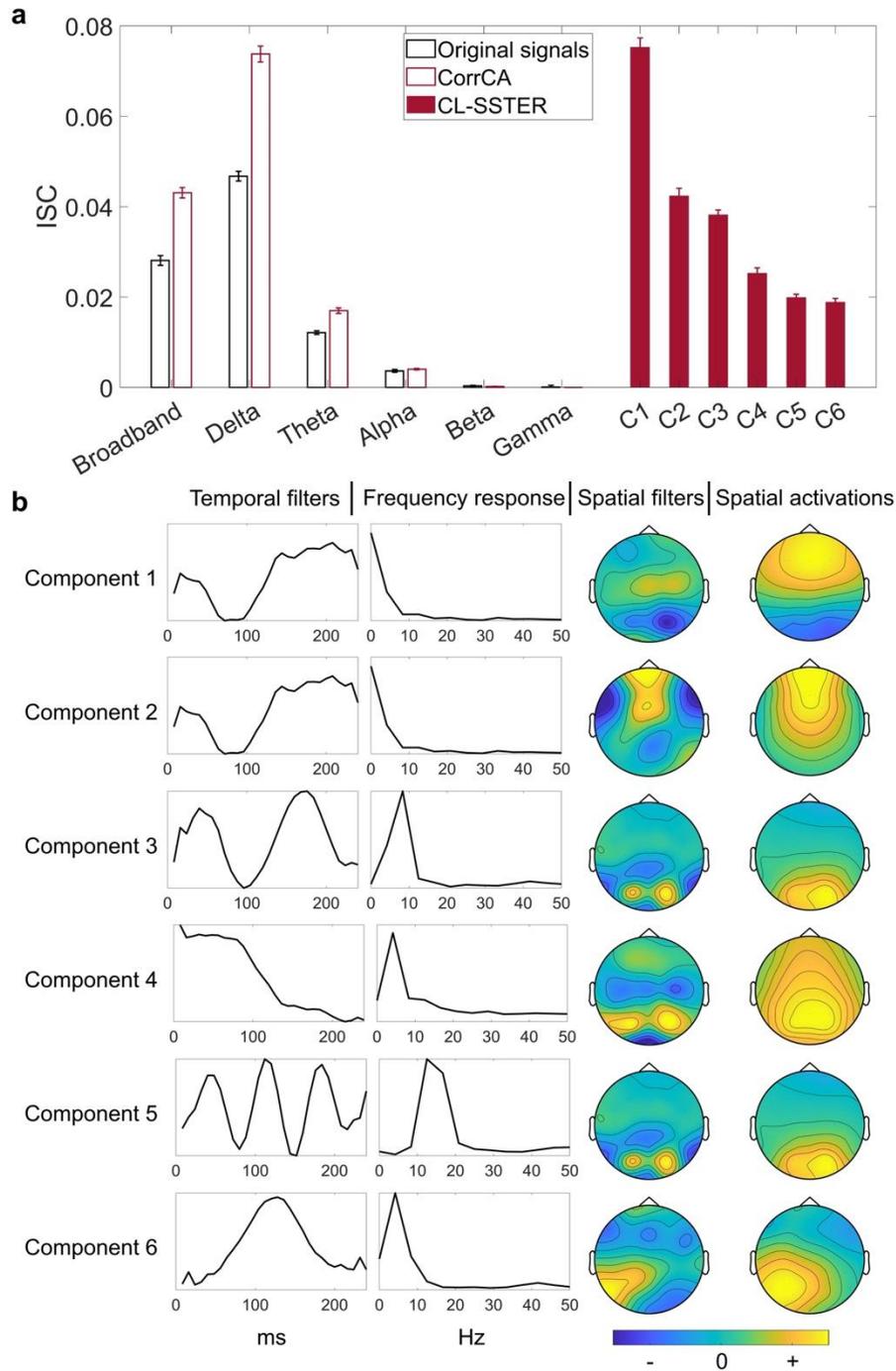

**Fig. 8 | ISCs of inter-subject shared components and their spatiotemporal characteristics on the FACED dataset. a** ISCs of different methods on the FACED dataset. The maximal ISCs of the original EEG signals and of the CorrCA components in each frequency band were shown as the left hollow bars. ISCs of the 6 components identified in CL-SSTER were shown as the right filled bars. "C1" to "C6" represent Component 1 to Component 6. **b** Visualization of representative temporal and spatial filters learned by CL-SSTER on the FACED dataset.

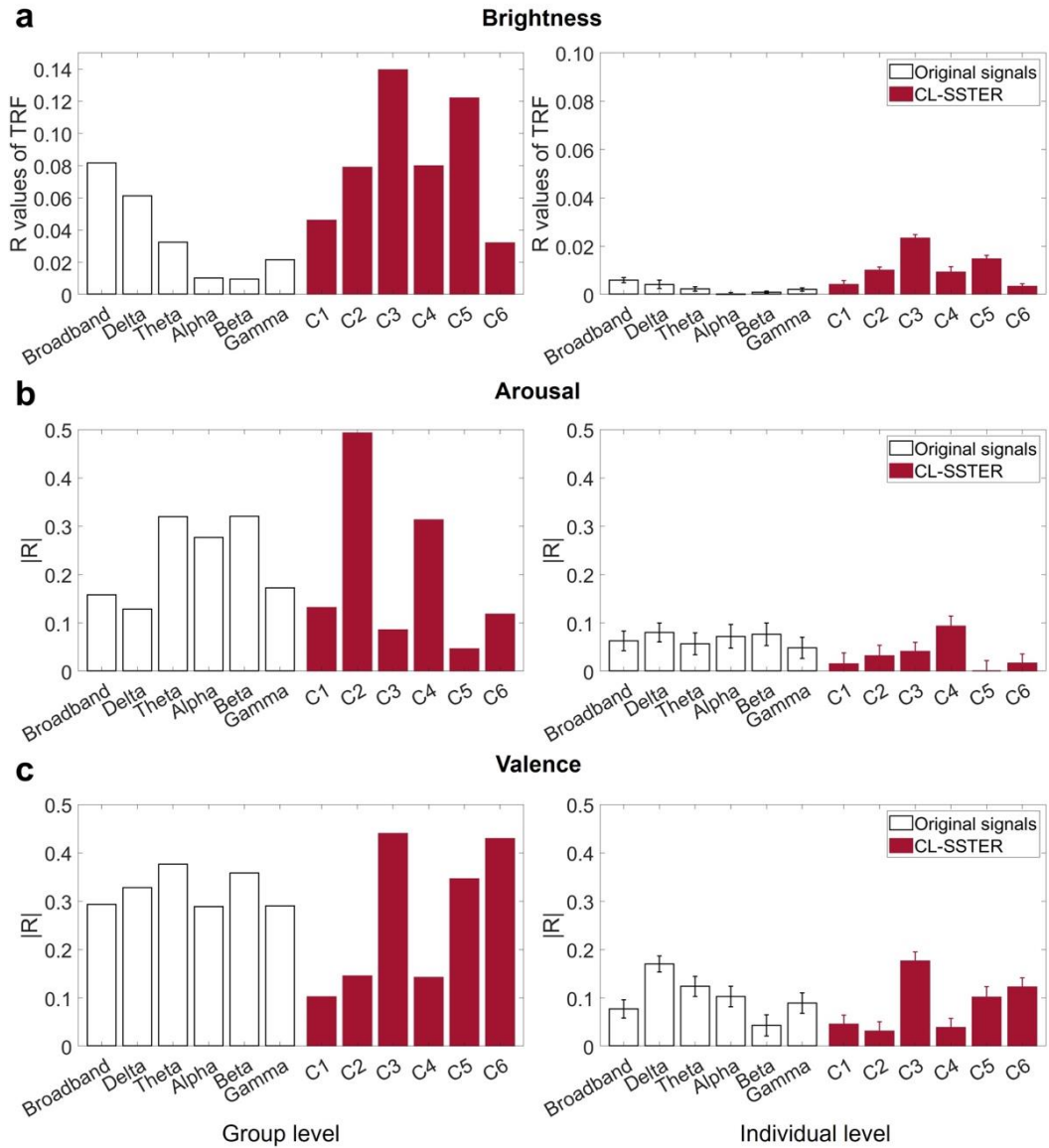

**Fig. 9 | The relationship between the components learned by CL-SSTER and the video features on the FACED dataset. a** R values of TRF models in predicting EEG signals or learned representations of CL-SSTER from video brightness. **b** Absolute correlations between EEG features and arousal ratings. **c** Absolute correlations between EEG features and valence ratings. Left column: group level. Right column: individual level. "C1" to "C6" represent Component 1 to Component 6.

# Supplementary information

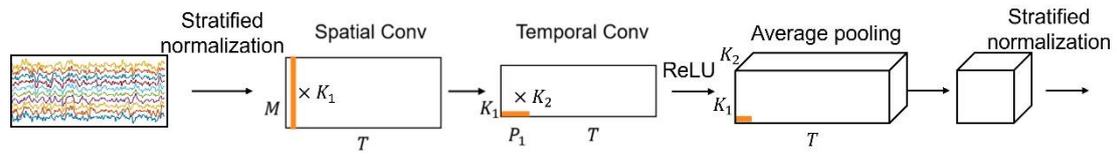

**Supplementary Figure 1. Details of the convolutional neural network architecture.** "Conv" means convolution in the figure.

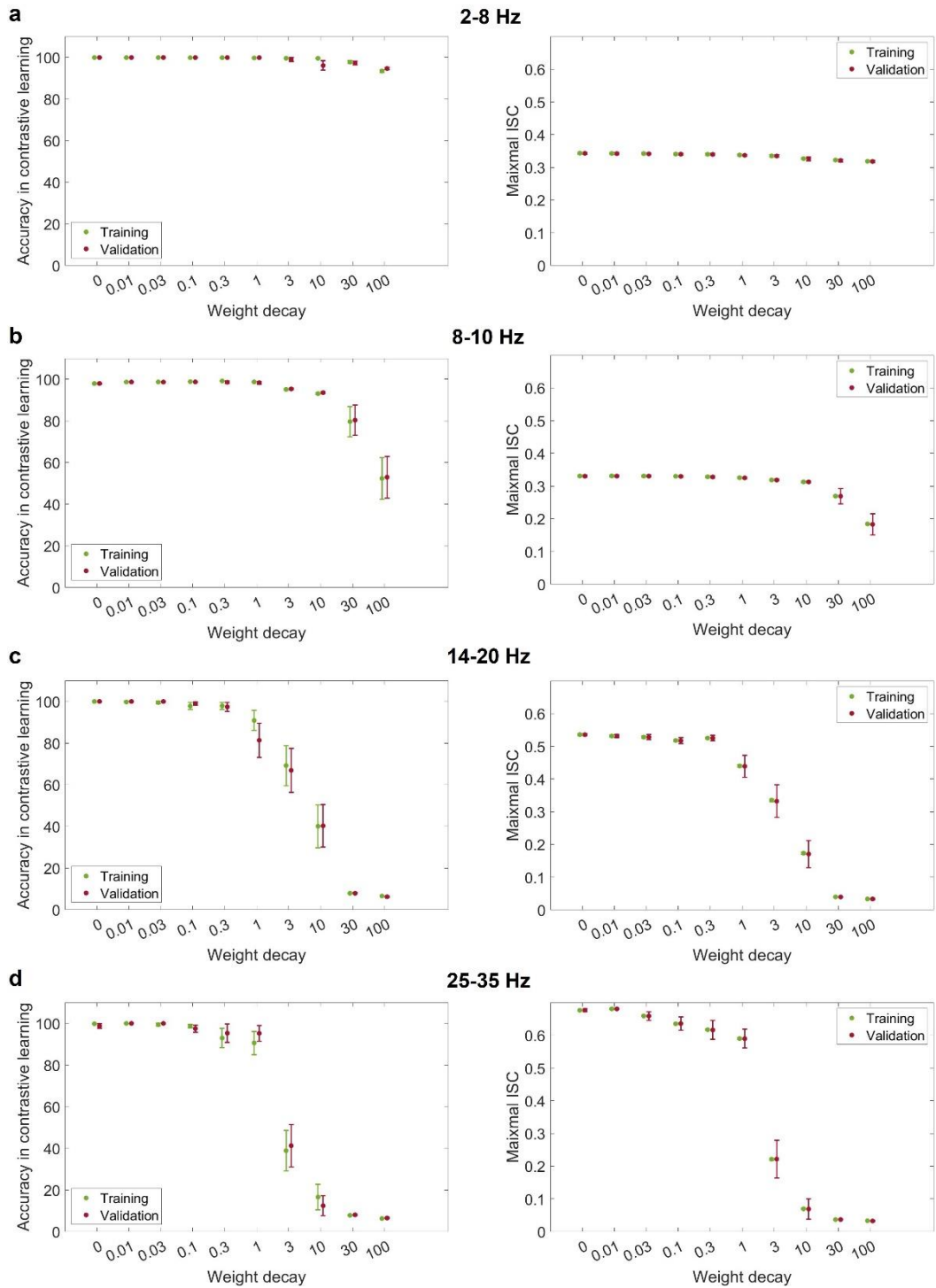

**Supplementary Figure 2. The effects of varying weight decay on model performance in the synthetic dataset.** The inter-subject shared oscillatory frequency ranges are 2-8 Hz (a), 8-10 Hz (b), 14-20 Hz (c) and 25-35 Hz (d), respectively. Left column: accuracy in contrastive learning. Right column: maximal inter-subject correlation (ISC).

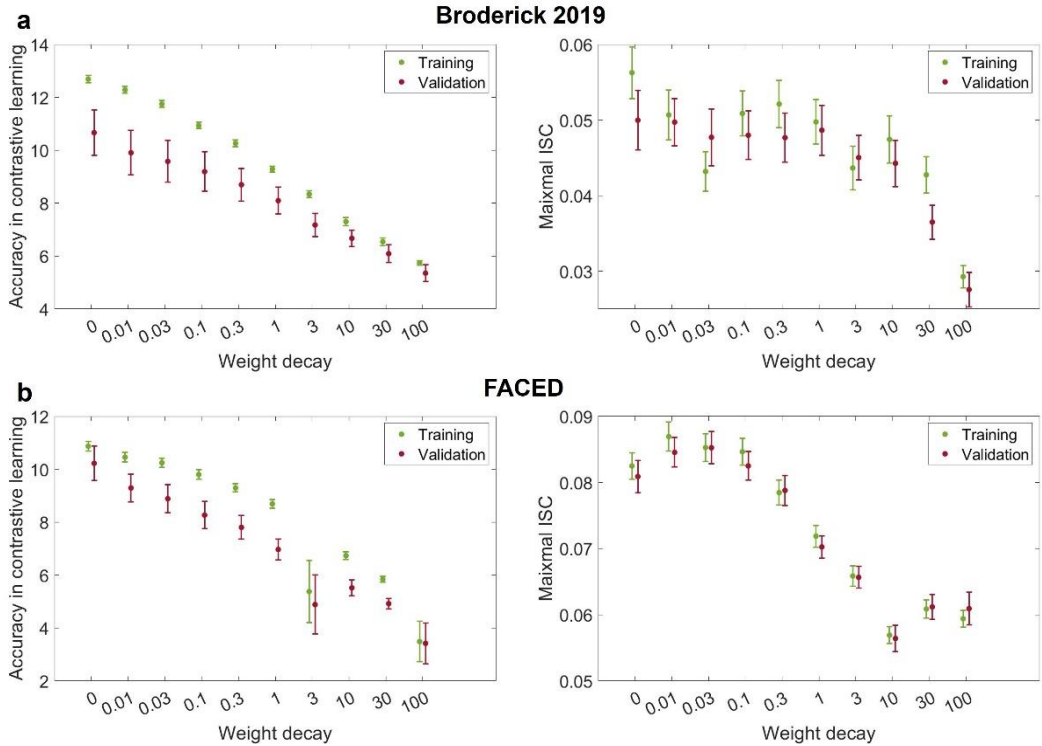

**Supplementary Figure 3. The effects of varying weight decay on model performance in the Broderick dataset (a) and the FACED dataset (b).** Left column: accuracy in contrastive learning. Right column: maximal inter-subject correlation (ISC).

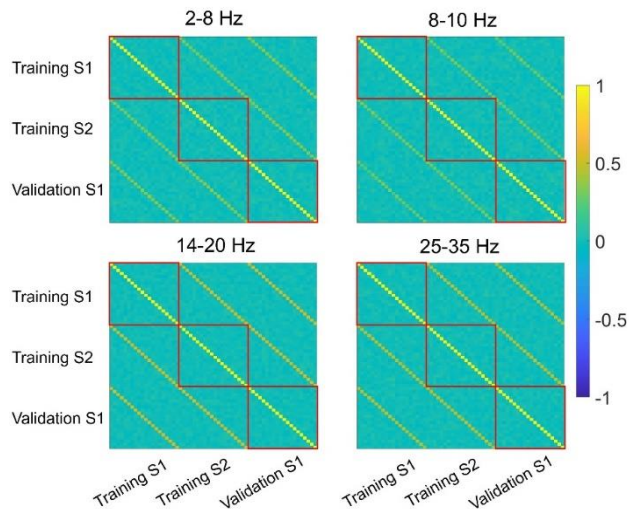

**Supplementary Figure 4. Cosine similarity of the representations across different trials on simulated data.** The results of 3 subjects (2 training subjects and 1 validation subject. S1 or S2 means subject 1 or subject 2) were shown in the figure. Each grid in the figure represents the cosine similarity between 2 trials. There were 20 trials for each subject (marked by a red rectangular). Auxiliary diagonals exhibit higher cosine similarity values, indicating representations of the corresponding trial pairs from different subjects are more closely aligned with greater similarity than other trial pairs.

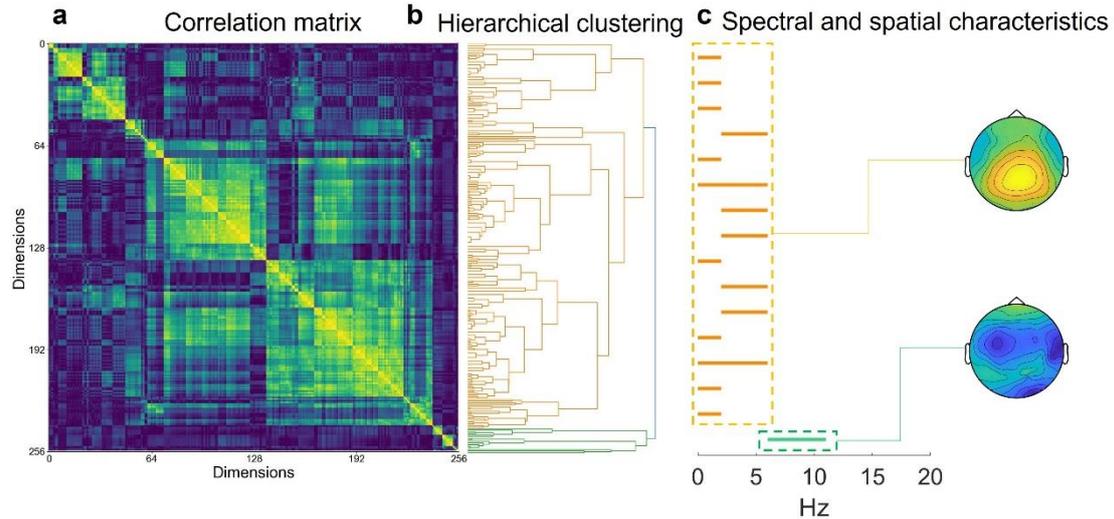

**Supplementary Figure 5. Clustering results on the Broderick dataset.** The correlation matrix of output dimensions (a) was submitted to hierarchical clustering, which identified two major components (b). The frequency ranges of the temporal filters in each cluster and spatial activations of the representative components were shown in (c). The positions of the left and right edges of the horizontal lines represent the lower and upper limits of the temporal filters' passband, respectively.

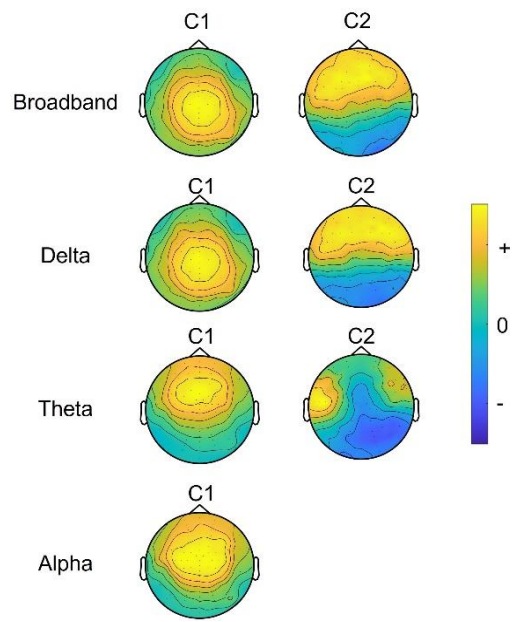

**Supplementary Figure 6. Visualization of spatial activations learned by CorrCA in each frequency band on the Broderick dataset.** For the broadband, delta and theta bands, two significant components were identified in each. One significant component was identified for the alpha band. No significant components were identified for the beta and gamma bands.

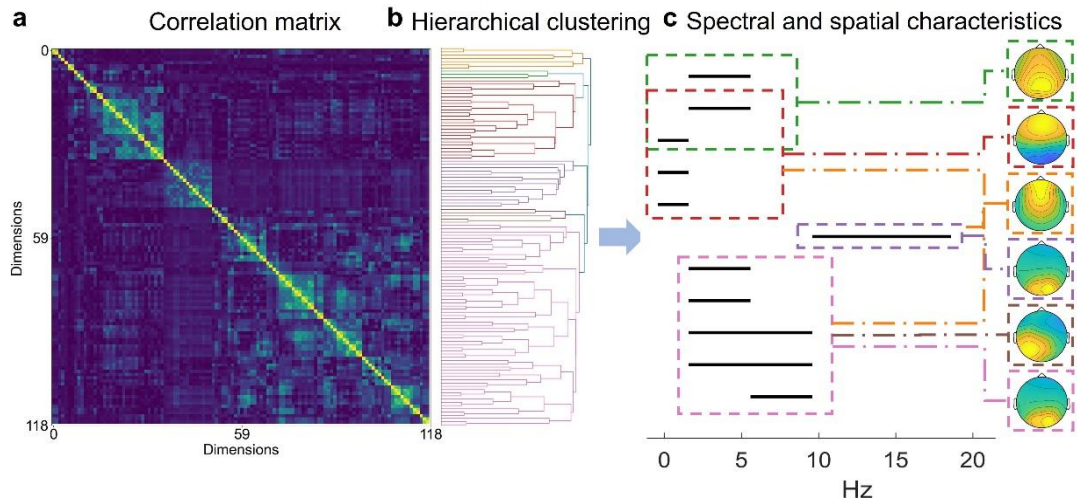

**Supplementary Figure 7. Clustering results on the FACED dataset.** Filters were clustered into six major components. The correlation matrix of output dimensions (a) was submitted to hierarchical clustering, which identified six major components (b). The frequency ranges of the temporal filters in each cluster and spatial activations of the representative components were shown in (c). The positions of the left and right edges of the horizontal lines represent the lower and upper limits of the temporal filters' passband, respectively.

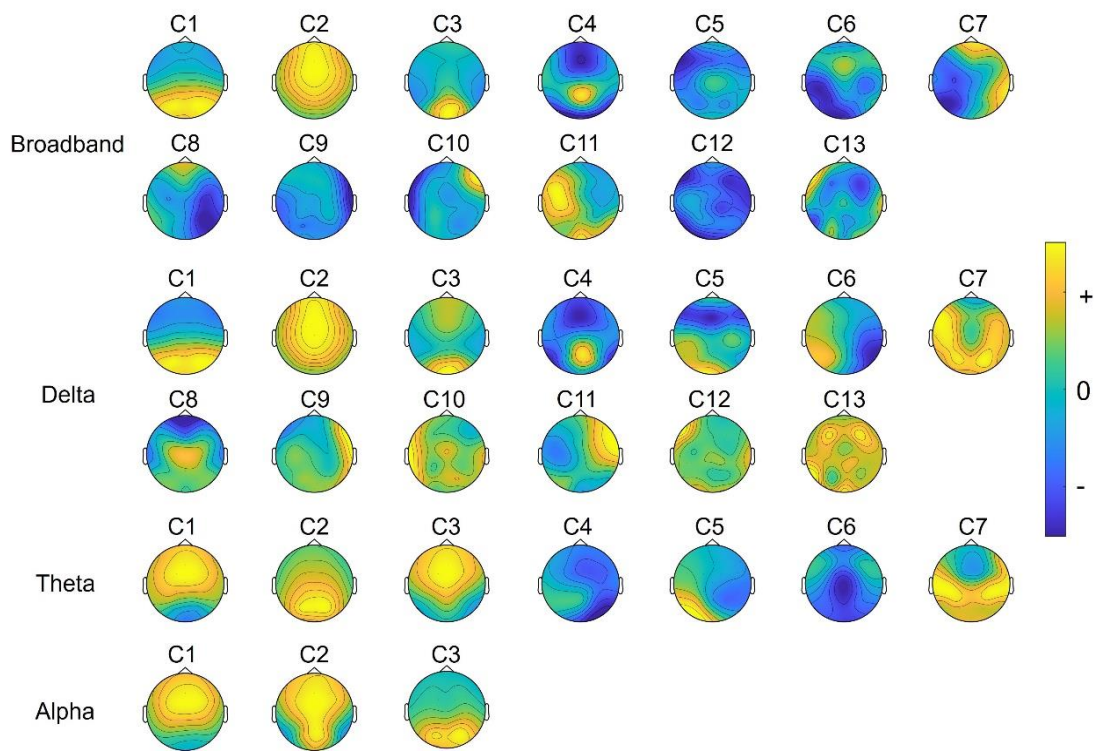

**Supplementary Figure 8. Visualization of spatial activations learned by CorrCA in each frequency band on the FACED dataset.** For the broadband and delta bands, 13 significant components were identified in each. Seven significant component was identified for the theta band. Three significant components were identified for the alpha band.